\documentclass[12pt]{article}
\usepackage{graphicx}
\usepackage{amsfonts}
\usepackage{amscd}
\usepackage{latexsym}
\usepackage{epsf}

\newcommand{\be}{\begin{equation}}
\newcommand{\ee}{\end{equation}}
\newcommand{\bea}{\begin{eqnarray}}
\newcommand{\eea}{\end{eqnarray}}

\newcommand{\fr}{\frac}

\newcommand{\al}{ \alpha }

\newcommand{\ep}{ \epsilon }

\newcommand{\ra}{\rightarrow}

\usepackage[latin9]{inputenc}
\usepackage{textcomp}
\usepackage{amsmath}
\usepackage{multirow}
\usepackage{color}
 
\newcommand {\si} {\sigma} 

\newcommand {\pb} {\bar{p}}

\begin{document}
\title{\bf Kinetics of Final Degassing of Hydrogen Desorption
 by Metal Hydrides}
\author{
{\sc I.~V.~Drozdov}\thanks{Corresponding author, e-mail: drosdow@uni-koblenz.de}
 }
\maketitle
\begin{center} 
\small Institute of Energy Research, 
\small Forschungszentrum J\"ulich \\
\end{center} 

\begin{abstract}
The proposed model concerns the 'confluent shrinking core' scenario and reproduces the
 desorption kinetic after the complete decay of the stoichiometric hydride ($\beta$-phase). The exact analytical solution is obtained, the numerical values are demonstrated by the example of magnesium hydride.
\end{abstract}

\small

\section{Introduction}

 The aim of the modelling outlined below was 
 to reproduce the final phase of the desorption in the case of complete outgassing of the 
desorbent. The maximal pressure $p_{max}$ reached through the desorption remains thereat below the threshold pressure $\pb=C_s X^2$ defined 
by the critical concentration of the $\al$ - dissolved hydrogen $X$ and the Sievert's constant $C_s$ for the given temperature, the volumetric {\bf case II}  \cite{1stpaper}.   
 
 The correct accounting of processes in the final desorption phase provides the observable shape of the kinetic curve. The main slope of the 'active desorption' merges smoothly into the asymptotic pressure $p=p_{max}$.
 This improves the main drawback of the simple model in the case II where the elbow  still remains at the top of desorption curve.  

 As it was found recently \cite{1stpaper, 2stpaper}
 the kinetics of hydrogen desorption can be described successfully by the ansatz
of 'confluent shrinking core' model \cite{gabis}. This approach allows to evaluate
 properly the kinetics of the pressure in the volumetric space increasing due the
desorption.

Following this scenario, the desorption rate controlling stage is the reversible 
process (chain of reactions) in the surface of the grain, responsible to the transition
of the hydrogen ions dissolved in the metallic lattice ($\al$-phase) into the molecular
 gas outside of the grain. In a linear approximation this process is represented by two
constants $b, k$, responsible for desorption and re-adsorption respectively.

The active {\it main phase} of desorption is the dew (dissolution) of the grain core consistent
of the $\beta$-hydride - the stoichiometric metal hydride with the spatial concentration of hydrogen atoms  $Y$ (for the $MgH_2\ \beta-core) $, is Y= 110119 $[mol/m^3]$. 
The main part of the desorption curve is defined by the kinetics of this phase, until the grain does not contain the stoichiometric $\beta$-hydride anymore. 
During this 'active desorption' the volume concentration $X$ of 
$\al$-dissolved hydrogen ions remains constant equal to the critical value for the
$\beta \ra \al$ transition (for magnesium this is typically X(T)=300-1500 $ [mol/m^3]$ for desorption, dependent on temperature T).

 Afterwards, the remaining $\al$-dissolved hydrogen leaves the grain via the same
Sievert's surface mechanism as before, up to the complete outgassing of the grain. This phase was named  'final degassing' \cite{gabis, castro}.
 It runs much more slowly as the previous 'active desorption',
 the total amount of hydrogen desorbed in the final degassing is 
\be
\nu_{fd}=\fr{\nu_{act} }{Y/(\eta X)-1},
\ee
about 94.34 times less compared to the active desorbed amount (for magnesium).

 The resulting increment of the pressure accounts after the main 'active desorption'
is about 1.06\%. Therefore the measurement requires a larger amount of desorbent to reach a higher pressure due the desorption. Otherwise, the final degassing phase is  very hard to distinguish experimentally because of natural fluctuations in the pressure during the measurement.
           
\subsection{ Quantitative Description}    

 All grains are assumed to be of an equal form and an equal size
 The start of the 'final degassing' phase relates to the moment, that the $\beta$
- stoichiometric metal hydride just decays completely whereby the pressure $p_0$
has been reached.
  
 The concentration $X(t)$ of $\al$-dissolved hydrogen decreases from the critical (maximal) value $X$ to zero.
\be
X(t)=X -\fr{\nu_i^{des}(t) }{\eta v}
\ee
  $\eta = 0.77$ is the volume shrinkage factor, $v$- the initial volume of a single $MgH_2$ grain,
 $\nu_i$ is the amount of desorbed hydrogen atoms $H$ [mol] from the single grain. 

The increase of the total pressure due the desorption from $N$ equal grains is

\be
p(t)= \fr{N  R}{ 2\{ V/T \} } \nu_i^{des}(t),
\label{pressure}
\ee
$R$-gas constant, $\{ V/T \}$ -effective ratio volume/temperature.

Using the hypothesis of the rate controlling step on the surface,
\be
\dot{ \nu}_i^{des}(t) = \fr{ 2\{  V/T \} }{N R} \dot{p}= 
s_i \left[      b\left(X - \fr{\nu_i^{des}(t) }{\eta v} \right)^2-k(p_0+p) \right]
\ee
and substituting the (\ref{pressure}) one arrives at
\be
\dot{p}=\fr{R\ N s_i}{2 \{ V/T \}  } \left[  
b\left(  X - \fr{2\{V/T\} p }{R\eta\ Nv_i^0} \right)^2-k(p_0+p)
\right],
\label{kinetic}
\ee
that is the {\it kinetic equation} for the final degassing. It have been assumed, 
the degassing starts as the $\beta-MgH_2$ decays completely and the pressure reaches the  
value $p_0= p_{\al\ra\beta}=\fr{N  R}{ 2\{ V/T \} } \nu_i^{act}$.

 The pressure $p$ means here the additional pressure arising from the final degassing
process in excess of the $p_0= p_{\al\ra\beta}$. With the notation used before
\be \ep:= \fr{m {\cal P} R }{2\{ V/T \}},\  m, {\cal P}\mbox{\small - mass and purity of the sample respectively,}\ee 
this equation is rewritten in an integrable form as
\be
\fr{d p}{ \fr{b}{k}\left( X-\fr{\varrho}{\eta\ep} p \right)^2- p_{\al\ra\beta} -p}=k\si\ep\eta^{2/3}\ d t=\xi\ d t,
\label{kinetic_integrable}
\ee
where $\xi:=k\si\ep\eta^{2/3}$ is introduced.

\subsection{Exact Solution}

Keeping in mind that $bX^2/k \equiv \pb $ is the threshold pressure, which is never reached
 (case II of \cite{1stpaper}), and using the notations:
\be \kappa = \fr{\varrho}{\eta\ep};\ \ \  \Delta:=\pb- p_{\tiny\al\ra\beta},\ \ \  C_s=b/k \ \mbox{ (Sievert's constant) }\ee
we obtain the result of the immediate integration of (\ref{kinetic_integrable}) 
\be
p(t)=p_{\al\ra\beta}+\fr{ p_2 - p_1 C e^{(p_2-p_1)\xi t}}{1 -  C e^{(p_2-p_1)\xi t}},
\ee
where $p(t)$ is now the total pressure in the volumetric system, $p_1$ and $p_2$ are two
real roots of the denominator in the left-hand side of (\ref{kinetic_integrable}), 
\be p_{1,2}= \fr{1}{2 C_s \kappa^2} \left[   (2 S X\kappa+1) \pm \sqrt{ (2 C_s X\kappa+1)^2-4\Delta C_s \kappa^2}\right]\ee
$C=p_2/p_1$- the constant of integration chosen to
obey the initial condition\\ $p(t=0) = p_{\al\ra\beta}$.

 An example modelled with the given initial parameters plotted in Fig.2 ( {\it Mathematica}), shows the slow asymptotical approach from $p_{\al\ra\beta}$ to $p_{max}$ as expected and reproduces the experimental results of \cite{meng} \\

\includegraphics[scale=0.66]{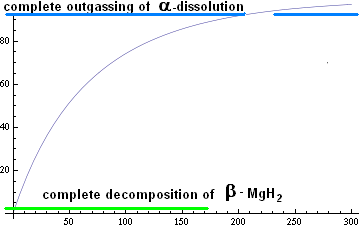}
{\bf\small
\hspace*{1cm}
\begin{minipage}[b]{6.5cm}
{\bf Fig.2 \\}
\mbox{Final desorption phase}\\
\mbox{for initial parameters:}\\
$
\si=75\cdot 10^3\\
\{V/T\}=1.007\cdot 10^{-7}\\
{\cal P}=0.8\\
m=20 \ mg\\
k=1.2\cdot 10^{-12}\\
C_s=0.5,\\
\mbox{providing}\\
\pb=180\ kPa,\\
p_{\al\ra\beta}=49.926\ kPa.\\
$
\end{minipage}
}\\

\section{Conclusion and Outlook}

 In the present paper the modelling of the hydrogen desorption kinetics developed 
in our previous works was extended by taking of the final degassing phase into account.
 Together with the surface shrinkage \cite{2stpaper}, which corrects the main slope of the active desorption,
the final degassing step models the merge to the asymptotic pressure in the case of complete outgassing.
This modeling does not require any additional parameters and is therefore a natural continuation of the previuosly considered models. The resulting solution is represented in a simple analytical form.

 The shape of the experimental desorption kinetics is reproduced therefore almost completely, excepting the 
initial delay phase, which is caused by local surface effects at the beginning, as suggested
in \cite{1stpaper, 2stpaper}. 
A quantitative understanding of this process is an important aspect in the practical application of the metal-alloy hydrogen storage and should be the subject of the future modeling activity.

\end{document}